\documentstyle[times,graphics,astrobib,amssymb]{mn2e}

\def\lsim{\mathrel{\rlap{\lower4pt\hbox{\hskip1pt$\sim$}}
    \raise1pt\hbox{$<$}}}
\def\gsim{\mathrel{\rlap{\lower4pt\hbox{\hskip1pt$\sim$}}
    \raise1pt\hbox{$>$}}}

\def\beq{\begin{equation}}
\def\eeq{\end{equation}}
\def\beqa{\begin{eqnarray}} 
\def\eeqa{\end{eqnarray}}

\def\laq{\raise 0.4 ex \hbox{$<$}\kern -0.8 em\lower 0.62 ex\hbox{$\sim$}}
\def\gaq{\raise 0.4 ex \hbox{$>$}\kern -0.7 em\lower 0.62 ex\hbox{$\sim$}}
\begin{document}
\title[GCG and recent Supernova data]
{Latest Supernova data in the framework of the Generalized Chaplygin Gas model}
\author[Bertolami et al.]
{O.~Bertolami$^1$\thanks{E-mail: orfeu@cosmos.ist.utl.pt},
A.A.~Sen$^1$\thanks{E-mail: anjan@cfif3.ist.utl.pt},
S.~Sen,$^2$\thanks{E-mail:somasri@cosmo.fis.fc.ul.pt},
and P.T.~Silva$^1$\thanks{E-mail:paptms@ist.utl.pt}\\
$^1$Instituto Superior T\'ecnico, Departamento de F\'\i sica, Av. Rovisco 
Pais, 1, 1049-001, Lisboa, Portugal \\
$^2$CAAUL, Departamento de F\'\i sica da FCUL, Campo Grande 1749-016, Lisboa, 
Portugal}
\maketitle

\date{\today}
\begin{abstract}

\noindent
We use the most recent Type-Ia Supernova data in order to 
study the dark energy - dark matter unification approach in the context of the 
Generalized Chaplygin Gas (GCG) model. Rather surprisingly, we find that data allow models with $\alpha > 1$.
We have studied how the GCG adjusts flat and non-flat models, and our results show that GCG is consistent with flat
case up to $68\%$ confidence level. Actually this holds even if one relaxes the flat prior assumption.
We have also 
analyzed what one should expect from a future experiment such as SNAP. We find that there is a degeneracy between
the GCG model and a XCDM model with a phantom-like dark energy component.
\end{abstract}

\section{Introduction}

Recent cosmological observations reveal that the Universe is dominated by two invisible components. Type-Ia Supernova 
observations \cite{Riess1998,Garnavich1998,Perlmutter1999},  nucleosynthesis constraints 
\cite{Burles2001}, Cosmic Microwave Background Radiation (CMBR) power spectrum \cite{Balbi2000,deBernardis2000,Jaffe2001}, 
large scale structure \cite{Peacock2001} and,
determinations of the matter density \cite{Bahcall1998,Carlberg1998,Turner2000} allow for a model where 
the clumpy 
component that traces matter, dark matter, amounts for about $23\%$ of the cosmic energy budget, while an overall 
smoothly distributed component, dark energy, amounts for approximately $73\%$ of the cosmic energy budget.

The most interesting feature of this dark energy component is that it has a negative pressure 
and drives the current accelerated expansion of the Universe \cite{Riess1998,Garnavich1998,Perlmutter1999}. 
From the theoretical side, great effort has been devoted to model dark energy.  The most obvious
candidate  is the vacuum energy, an uncanceled
cosmological constant [see eg. \citeN{Bento1999}, \citeN{Bento2001b}] for which $\omega_x \equiv p_x / \rho_x = -1$.
Another possibility is a dynamical vacuum  \cite{Bronstein1933,Bertolami1986a,Bertolami1986b,Ozer1987} or quintessence. 
Quintessence models most often involve a single
scalar field 
\cite{Ratra1988a,Ratra1988b,Wetterich1988,Caldwell1998,Ferreira1998,Zlatev1999,Binetruy1999,Kim1999,Uzan1999,Amendola1999,Albrecht2000,Bertolami2000,Banerjee2001a,Banerjee2001b,SenSen2001,Sen2001} or two coupled
fields  \cite{Fujii2000,Masiero2000,Bento2002a}. In these models, the cosmic coincidence problem, 
that is, why did the dark  energy start to dominate the cosmological evolution
only fairly recently, has no satisfactory solution and some fine tuning is required.

More recently, it has been proposed that the  evidence for a dark energy 
component might be explained by a change in the equation 
of state of the background fluid, with an exotic equation of state, the generalized Chaplygin gas (GCG) model,
rather than by  a cosmological constant 
or the dynamics of a scalar field rolling down a potential 
\cite{Kamenshchik2001,Bilic2002,Bento2002b}. In this proposal, 
one considers the evolution 
of the equation of state of the background fluid instead of a quintessence
potential.
The striking feature of this model is that it  allows for an
unification of dark energy and dark matter
\shortcite{Bento2002b}. Moreover, it is shown that the GCG model
may be accommodated within the standard structure formation scenario
\shortcite{Bilic2002,Bento2002b}. Concerns about this point have been raised 
by \citeN{Sandvik2002}, however in this analysis, the effect of baryons has not been taken into account, 
which was shown to be important and allowing compatibility with the 2DF mass
power spectrum \cite{Beca2003}. Also, the \citeN{Sandvik2002} claim was based on the
linear treatment of perturbations close to the present time, thus neglecting
any non-linear effects.

Thus, given it potentialities, the GCG model has been the subject
of great interest, and various attempts have been made to constrain its parameters
using the available observational data. Studies include Supernova data and power spectrum \cite{Avelino}, 
age of the Universe and strong lensing statistics \cite{Dev1}, age of the Universe and Supernova data 
\cite{Makler,Alcaniz}. The tightest
constraints were obtained by \citeN{Bento2003a} using the CMBR power spectrum
measurements from BOOMERANG \cite{Boomerang} and Archeops \cite{Archeops},
together with  the SNe Ia constraints. It is shown that
$0.74 \lsim A_s \lsim 0.85$, and $\alpha\lsim 0.6$, ruling out the pure
Chaplygin gas model. From the bound arising from the age of the APM 08279+5255
source, which is $A_s\gsim 0.81$ \cite{Alcaniz}, one can get tight
constraints, namely  $0.81 \lsim A_s \lsim 0.85$, and
$0.2 \lsim \alpha\lsim 0.6$, which also rules out the $\Lambda$CDM model.
These results were in agreement with the WMAP data \cite{Bento2003b}. It was
also shown that the gravitational lensing statistics from future large surveys
together with SN Ia data from SNAP will be able to place interesting
constraints the parameters of GCG model \cite{Silva2003}. As we shall see in Sections 3 and 4, all these constraints 
are consistent with Supernova data at $95\%$ confidence level.

Recently \citeN{Choudhury2003b} have analyzed the supernova data 
with currently available 194 data points [see also \citeN{Choudhury2003a}] and shown that it 
yields relevant constraints on some cosmological parameters. In particular, it shows that when one considers 
the full supernova data set, it rules out the decelerating model with significant confidence 
level. They have also shown that one can measure the current value of the dark energy equation 
of state with higher accuracy and the data prefers the phantom kind of equation of state, 
$\omega_X < -1$ \cite{Caldwell2002}. Moreover, the most significant  observation of their analysis is that,
without a flat prior, the latest Supernova data also rules out the preferred flat $\Lambda$CDM  model which 
is consistent with other cosmological observations. 
In a previous paper, \citeN{Alam2003} have reconstructed the equation of state 
of the dark energy component using the same set of Supernova data and found that the dark 
energy evolves rapidly from  $\omega_x \simeq 0$ in the past to a strongly negative equation 
of state ($\omega_x \lsim -1$) in the present, suggesting that $\Lambda$CDM may not be a good choice for dark energy.
More recently, other groups  have also analyzed these recent
Supernova data in the context of different cosmological models for dark energy \cite{Gong2004,Perivol2004}. 

In this paper, we analyze the GCG model in the light of the latest supernova 
data \cite{Tonry2003,Barris2003}. We consider both  flat and 
non-flat models. Our analysis shows that the problem with the flat model, which has been discussed in \citeN{Choudhury2003b}, can be solved in the GCG model in
 a sense that flat GCG model is consistent with the latest Supernova data even without a flat prior.We have also 
analyzed the confidence contours for a GCG model, that one expects from a future experiment such as SNAP. 
We find that there is a degeneracy between
the GCG model and a XCDM model with a phantom-like dark energy component.

This paper is organized as follows. In Section 2 we discuss 
various aspects of the generalized Chaplygin gas model and its theoretical 
underlying assumptions.
In Section 3 we describe our best fit analysis of the most recent supernova 
data in the context of generalized Chaplygin gas model. Section 4 contains 
our analysis for expected SNAP results. Finally, in Section 5 we present 
our conclusions.


\section{Generalized Chaplygin Gas model}

The generalized Chaplygin gas (GCG) is characterized by the equation of state

\beq
p_{ch} = - {A \over \rho_{ch}^\alpha},
\eeq
where $A$ and $\alpha$ are positive constant. For $\alpha=1$ the equation of state is 
reduced to so-called Chaplygin gas scenario first studied in cosmological context by 
\shortciteN{Kamenshchik2001}. Inserting the above equation of state in the 
energy conservation equation, one can integrate it to obtain \shortcite{Bento2002b}

\beq
\rho_{ch} = \rho_{ch0} \left(A_{s} + {(1-A_s) \over a^{3(1+\alpha)}}\right)^{1/(1+\alpha)},
\eeq
where $\rho_{ch0}$ is the present energy density of GCG and $A_s \equiv 
A/\rho_{ch0}^{(1+\alpha)}$.

One of the most striking features of this expression is that, the energy density of this GCG, 
$\rho_{ch}$, interpolates between a dust dominated phase, $\rho_{ch} \propto a^{-3}$, in the 
past and a de-Sitter phase, $\rho_{ch} = -p_{ch}$, at late times. This property
makes the GCG model an interesting candidate for the unification of dark matter and dark energy. 
Indeed, it can be shown that the GCG model admits inhomogeneities and that, 
under the Zeldovich approximation, they evolve in a qualitatively similar fashion  like the $\Lambda$CDM model \cite{Bento2002b}. Furthermore, this evolution is controlled by the 
homogeneous parameters of the model, namely, $\alpha$ and $A$.

There are several important aspects of the above equation which one should discuss before 
constraining the relevant parameters using Supernova data. Firstly, one can see from the above 
equation that $A_s$ must lie in the range $0\le A_s \le 1$. For $A_s =0$, GCG behaves 
always as matter whereas for $A_s =1$, it behaves always as a cosmological constant. Hence 
to use it as a unified candidate for dark matter and dark energy one has to exclude these 
two possibilities resulting the range for $A_s$ as $0< A_s < 1$. 

To have an idea about the possible range for $\alpha$, one has to consider the propagation 
of sound through this fluid. Given any Lagrangian ${\cal L}(X,\phi)$ for a field $\phi$, 
where $X = {\frac{1}{2}}g^{\mu\nu}\phi_{,\mu}\phi_{,\nu}$, the effective speed of sound 
entering the equations for the evolution of small fluctuations is given by 
\beq
c_{s}^{2} = {p_{,X} \over \rho_{,X}} = {{\cal L}_{,X} \over {\cal L}_{,X}+2X {\cal L}_{,XX}}.
\eeq
Thus, for a standard scalar field model which has canonical kinetic energy term like
${\cal L} = X - V(\phi)$, the speed of sound is always equal to 1  irrespective of the equation of state. 
But for a Lagrangian containing a non-canonical kinetic energy term, one can have a sound speed 
quite different from 1. Actually, even $c_{s}^{2} > 1$ is possible which physically means
that the perturbations of the background fluid can travel faster than light as measure
in the preferred frame where the background is homogeneous. For a time dependent
background field, it does not lead to any violation of causality as the underlying
theory is manifestly Lorentz invariant \cite{Erickson2002}. 

For GCG it has been shown that the equation of state (1) can be obtained from a generalized version of the 
Born-Infeld action \shortcite{Bento2002b}
\beq
{\cal L} = -A^{1/(1+\alpha)}\left[1 - (g^{\mu\nu}\phi_{,\mu}\phi_{,\nu})^{(1+\alpha)/2\alpha}\right]^{\alpha/(1+\alpha)}~~,
\eeq
which for $\alpha = 1$ leads to the Born-Infeld action. If one computes $c_{s}^2$ for this action 
one can get using Eq. (2) the present value as $c_{s0}^2 = \alpha A_{s}$. As $A_{s}$ is always positive, its 
restricts $\alpha$ to only positive values. In all previous work, $\alpha$ has been restricted to 
a value up to 1. But one can see from the above expression for $c_{s0}^2$ that as $0 < A_s < 1$, 
the maximum allowed value for $\alpha$ can be surely greater than 1 and that also depends on the 
value of $A_s$, e.g for $A_s = 0.5$, the allowed range for $\alpha$ is $0 \leq \alpha \leq 2$. 
Notice also that the dominant energy condition $\rho+p \ge 0$ is always valid in this case. Furthermore, there is 
no big rip in the future and asymptotically the Universe goes toward a de-Sitter phase. 

Hence, on general grounds, restricting $\alpha$ up to 1 is not a very justified assumption. Moreover,
this restriction  arises mainly by considering the present day value of $c_{s}^2$ which is not an
important epoch for structure formation. In general $c_{s}^2$ in this model is a time dependent
quantity and in such  cases it is not very proper to constrain $\alpha$ with the present day value of $c_{s}^2$. 

There is another reason for not restricting $\alpha$ up to 1. It has been shown by \shortciteN{Kamenshchik2001} 
that one can also model the Chaplygin gas with a minimally coupled
scalar field with canonical kinetic energy term in the Lagrangian density. Performing this
exercise for the GCG, leads to a potential for this scalar field of the form
\beq
V = V_{0}e^{3(\alpha-1)\phi}[\cosh({m\phi\over{2}})^{2/(\alpha+1)}
 + \cosh({m\phi\over{2}})^{-2\alpha/(\alpha+1)}]
\eeq
where $V_{0}$ is a constant and $m = 3(\alpha+1)$. For $\alpha=1$, one recovers the potential 
obtained by \shortciteN{Kamenshchik2001}. Now as we have discussed earlier, for a minimally coupled scalar 
field with a canonical kinetic energy term, the value of $c_{s}^2$ is always 1 irrespective of the 
equation of state. Hence if one considers this kind of scalar field to model GCG, there is no such 
restriction on $\alpha$ coming from the sound speed.

\begin{figure}
\begin{center}
{\resizebox{0.45\textwidth}{!}{\includegraphics{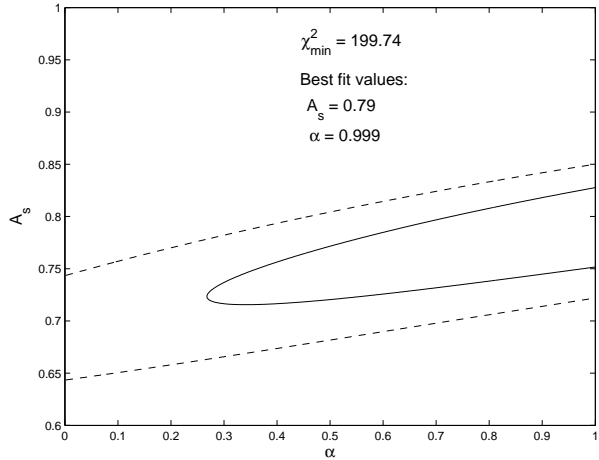}}}
\caption{Confidence contours in the $\alpha - A_s$ parameter space for flat unified GCG model. The 
solid and dashed lines represent the $68\%$ and $95\%$ confidence regions, respectively. The best fit 
value used for ${\cal M}^{'}$ is -0.033.
}
\label{figure1}
\end{center}
\end{figure}

\begin{figure}
\begin{center}
{\resizebox{0.45\textwidth}{!}{\includegraphics{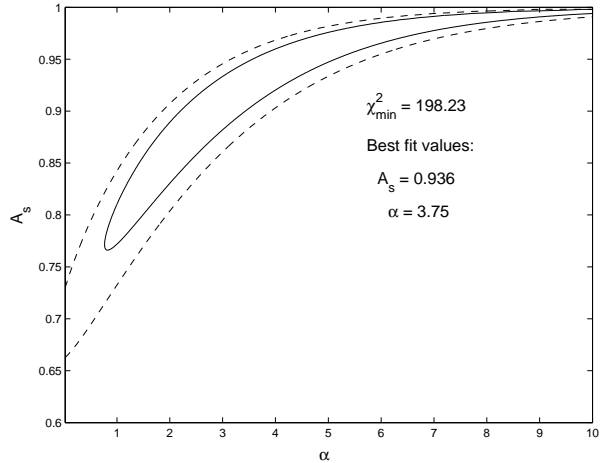}}}
\caption{Same as Figure 1, but with a wider range for $\alpha$.
}
\label{figure2}
\end{center}
\end{figure}

In what follows, we shall consider that $A_s$ lies in the range $0 <  A_{s} < 1$ and the 
only constrain on $\alpha$ that we shall consider is that it takes positive values. We should 
also point out that the $\alpha = 0$ case corresponds to the $\Lambda$CDM model.

The Friedmann equation for a non-flat unified GCG model in general  is given by
\begin{eqnarray}
H^{2} = H_{0}^{2}\hspace{2mm}[\Omega_{ch}\left(A_{s} + 
(1-A_s)(1+z)^{3(1+\alpha)}\right)^{1/(1+\alpha)} \nonumber\\
 + \hspace{2mm}\Omega_{k}(1+z)^{2}]~~,
\end{eqnarray}
where $H_{0}$ is the present day value of the Hubble constant. $\Omega_{ch}$ and $\Omega_{k}$ are the 
present day density parameters for GCG and the curvature. For a flat Universe $\Omega_{k}=0$ 
and $\Omega_{ch}=1$, whereas for the non-flat case $\Omega_{ch}=1-\Omega_{k}$.


\section{Recent Supernova data and the best fit analysis}


To perform the best fit analysis of our GCG model with the recent Supernova data, we follow the 
method discussed by \citeN{Choudhury2003b}. As far as 
the Supernova observation is concerned, the cosmologically relevant quantity is the 
{\it apparent magnitude}, $m$, given by

\beq
m(z) = {\cal M} + 5 \log_{10} D_{L}(z)~~,
\eeq
where $D_{L} = {\frac{H_{0}}{c}}d_{L}(z)$, is the dimensionless luminosity distance, and the 
luminosity distance $d_{L}(z)$ is given by $d_{L}(z) = r(z)(1+z)$ where $r(z)$ is the comoving 
distance. Also ${\cal M} = M + 5 \log_{10} \left(\frac{c/H_0}{1~Mpc}\right) + 25$ where 
$M$ is the absolute magnitude for the Supernova which is believed to be constant for all Type-Ia Supernova.

In our analysis, we take the 230 data points listed in \citeN{Tonry2003} along with 
the more recent 23 points from \citeN{Barris2003}. Also, as discussed by \citeN{Choudhury2003b}, for low redshifts,
data might be affected by the peculiar 
motions, making the measurements of the cosmological redshifts uncertain; hence we shall consider 
only those points with redshifts $z > 0.01$. Moreover, since it is difficult to be sure about the host galaxy 
extinction, $A_{v}$, we do not consider points which have $A_{v} > 0.5$. Hence in our final 
analysis, we consider only 194 points, which are similar to those
considered by \citeN{Choudhury2003b}.

The Supernova data points given by \citeN{Tonry2003} and \citeN{Barris2003} 
are listed in terms of luminosity distance $\log_{10}d_{L}(z)$ together with the corresponding error 
$\sigma_{\log_{10}d_{L}}(z)$. These distances are obtained assuming some value of $\cal{M}$ which may 
not be the true value. Hence, in our analysis we shall keep it as free parameter while fitting the data.

The best fit model is obtained by minimizing the quantity

\beq
\chi^2 = \sum_{i=1}^{194} \left[
{\log_{10}d_{L\rm obs}(z_i) 
- 0.2 {\cal M}' 
- \log_{10}d_{L\rm th}(z_i; c_{\alpha}) \over \sigma_{\log_{10}d_{L}}(z_i)}
\right]^2
\label{chisq}
\eeq
where ${\cal M}^{'} = {\cal M} - {\cal M}_{obs}$ is a free parameter denoting the difference between 
the actual ${\cal M}$ and the assumed value ${\cal M}_{obs}$ in the data. As discussed by \citeN{Choudhury2003b},
we have also taken into account the uncertainty arising 
because of the peculiar motion at low redshift by adding an uncertainty of 
$\Delta v = 500$ km s$^{-1}$ to $\sigma_{\log_{10}d_{L}}^2(z)$,
\beq
\sigma_{\log_{10}d_{L}}^2(z) \to 
\sigma_{\log_{10}d_{L}}^2(z) + \left({1 \over \ln 10}{1 \over d_{L}}{\Delta v \over c}\right)^2~~.
\eeq
This correction is more effective at low redshifts, i.e for small values of $d_{L}$. 

In our subsequent best fit analysis, the minimization of (\ref{chisq}) is done with respect to ${\cal M}^{'}, 
\alpha, A_s$ and $\Omega_{k}$. The parameter ${\cal M}^{'}$ is a model independent parameter and 
hence its best fit value should not depend on a specific model. We have checked that when 
minimizing (\ref{chisq}) with respect to ${\cal M}^{'}$, the best fit value for ${\cal M}^{'}$ for 
all of the models considered here is $-0.033$ which is also consistent with that obtained 
by \citeN{Choudhury2003b}. Hence, in our 
subsequent analysis, we shall use always this best fit value, ${\cal M}^{'} = -0.033$. 


\subsection{Flat case}

For this case, we assume $\Omega_{k} = 0$ and consider only two parameters, $\alpha$ and $A_s$. 
We first restrict $\alpha$ to be $\leq1$. In Figure 1, we have shown the 
$68\%$ and $95\%$ confidence contours in $\alpha - A_s$ parameter space. The best fit values for [$\alpha$, $A_s$]
 are given by [0.999, 0.79]. The best fit value of $\alpha$ is very close to its upper limit 
since the  actual best fit value lies in the region beyond $\alpha = 1$. Also, up to 
$68\%$ confidence level, the $\alpha = 0$, i.e. the $\Lambda$CDM case, is excluded although 
it is consistent at $95\%$ confidence level.

\begin{figure}
\begin{center}
{\resizebox{0.45\textwidth}{!}{\includegraphics{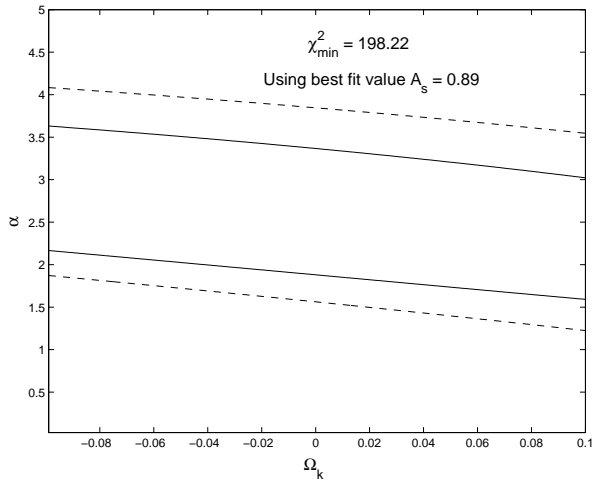}}}
\caption{Confidence contours in the $\Omega_k - \alpha$ parameter space for non flat unified GCG model. 
The solid and dashed lines represent the $68\%$ and $95\%$ confidence regions, respectively. The best 
fit value used for ${\cal M}^{'}$ is -0.033.
}
\label{figure3}
\end{center}
\end{figure}

\begin{figure}
\begin{center}
{\resizebox{0.45\textwidth}{!}{\includegraphics{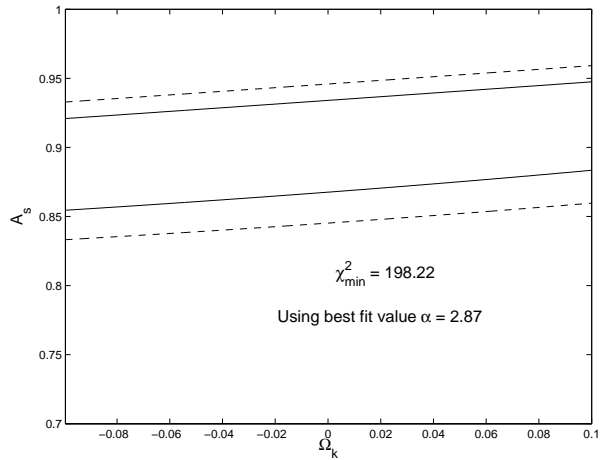}}}
\caption{Same as Figure 3, but in the $\Omega_k - A_s$ parameter space.
}
\label{figure4}
\end{center}
\end{figure}

Next we allow $\alpha$ to vary for a wider range. The Figure 2 represents the same as Figure 1 but with 
$\alpha$ taking a wider range. The best fit values for [$\alpha, A_s$] are  now [3.75, 0.936]. 
This high value of $\alpha$ might just be a statistical fluke though, as the confidence 
regions exhibit a very shallow valley along the $\alpha$ direction.
Here, also at $68\%$ confidence level, the 
$\alpha=0$, $\Lambda$CDM case, is excluded, although 
it is consistent at $95\%$ confidence level. It should be noted that both Choudhury \& Padmanabhan (2003)
and Tonry et al. (2003) also found some conflict between the $\Lambda$CDM model and the SN Ia data,
namely, that when imposing a flat Universe prior, the data ruled out the vacuum energy as an allowed 
dark energy component, and actually favoured a phantom energy component. Here we see that
the GCG model fits the data well, and, as mentioned previously, without the theoretical
complications that plague the phantom energy model, namely the dominant energy condition
is not broken, and there is no big rip singularity in the future. 

It is also clear from the minimum value for $\chi^2$ obtained in these two cases, that when one 
allows to vary $\alpha$ beyond $1$, one obtains a better fit to the Supernova data.


\subsection{Non-flat case}

Another problem that the $\Lambda$CDM has with the new SN Ia data is that without a flat prior, a flat $\Lambda$CDM  
Universe has also been ruled out at 68\% confidence \cite{Choudhury2003b}. \citeN{Tonry2003} argued
that this was probably due to some overlooked systematic error, since a small systematic error of $0.04$ mag 
was able make the flat $\Lambda$CDM consistent with the data. Here we attempt to find
if the GCG model might alleviate this problem.

For this, we allow a non-vanishing curvature in our model. We now have three parameters in our model namely, 
$\alpha$, $A_s$ and the density parameter for the curvature at present, $\Omega_k$. 
First we assume that our Universe deviates slightly from the flat model assuming $\Omega_k$ to 
vary between [-0.1,0.1]. In this case the best fit values for $\alpha, A_s$ and $\Omega_k$ 
are [2.87, 0.89, -0.099]. It suggests that the data prefers a negative 
curvature.
In Figure 3, we have shown the $68\%$ and $95\%$ confidence contours in the $\Omega_k$-$\alpha$ 
plane assuming the best fit value for $A_s$, whereas in Figure 4, we have shown the same contours 
in the $\Omega_k$-$A_s$ plane assuming the best fit value for $\alpha$.

\begin{figure*}
\begin{center}
{\resizebox{1\textwidth}{!}{\includegraphics{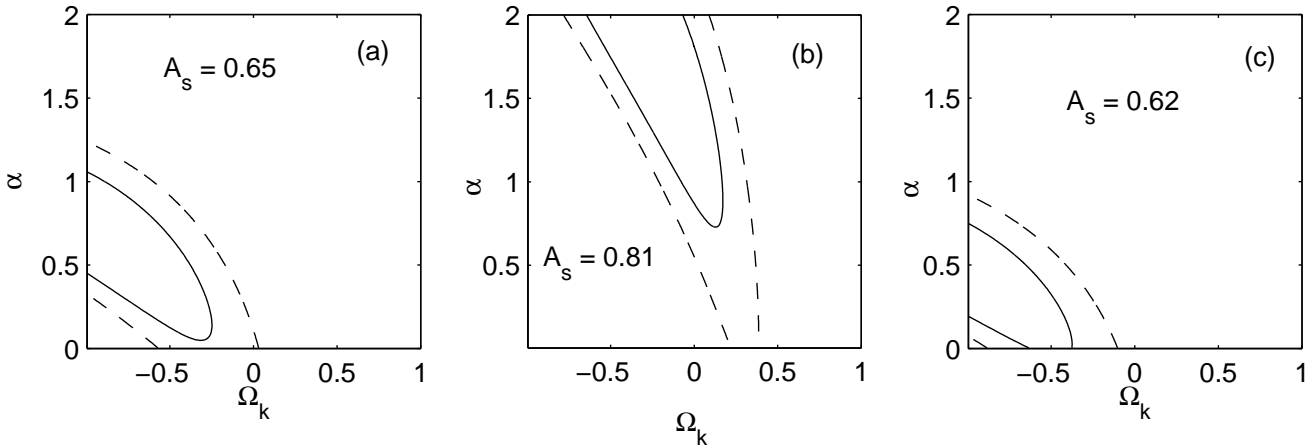}}}
\caption{Confidence contours in the $\Omega_k - \alpha$ parameter space for non-flat 
unified GCG model. The solid and dashed lines represent the $68\%$ and $95\%$ confidence 
regions, respectively. The best fit value used for ${\cal M}^{'}$ is -0.033. Figure (a) is for the best fit mean value
of $A_s$, whereas (b) and (c) are for the values in the wings of $68\%$ confidence limit. 
}
\label{figure5}
\end{center}
\end{figure*}
\begin{figure*}
\begin{center}
{\resizebox{1\textwidth}{!}{\includegraphics{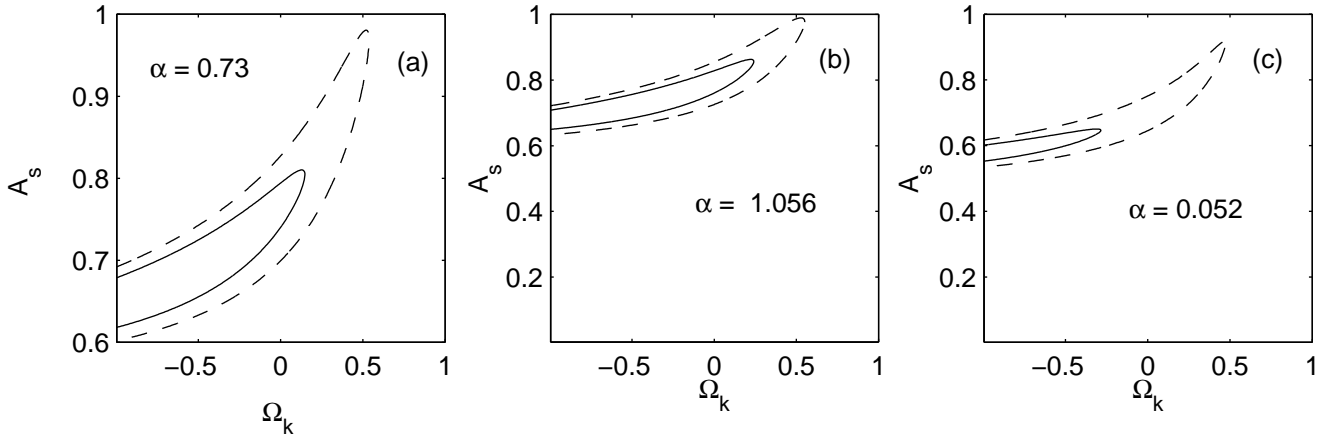}}}
\caption{Same as Figure 5, but in the $\Omega_k - A_s$ parameter space. Figure (a) is for the best fit value of
$\alpha$ whereas (b) and (c) are for the values in the wings of $68\%$ confidence limit.
}
\label{figure6}
\end{center}
\end{figure*}

Both $\alpha$ and $A_s$ are constrained significantly and $68\%$ confidence limits on $\alpha$ and $A_s$
are [1.6 3.625)] and [0.856 0.946)] respectively.
It shows that the data prefers higher value for $\alpha$ and the $\Lambda$CDM 
limit ($\alpha = 0$) is excluded, both for $68\%$ and $95\%$ confidence limit. Also one 
can see from both figures that the flat case $\Omega_k = 0$ is consistent with the data for 
both $68\%$ and $95\%$ confidence level, but for a higher value of $\alpha$.

We allow now more curvature in our model and consider the range for $\Omega_k$ to vary as [-1, 1]. 
In this case the best fit values for $\alpha, A_s$ and $\Omega_k$ are [0.73, 0.65, -0.999] and 
the resulting $\chi^2_{min}$ is 197.99 showing that there is an improvement in the quality of fit.
The $68\%$ confidence limits on $\alpha$ and $A_s$ are [0.052 1.056] and [0.62 0.81], respectively.
 Also the allowed range for $\alpha$ shifts more significantly towards smaller values and data do
allow the model to get closer to $\Lambda$CDM ($\alpha = 0.052$) as well as to the Chaplygin gas
model ($\alpha = 1$).

In Figure 5, we have shown the $68\%$ and $95\%$ confidence contours in $\Omega_k$-$\alpha$ 
plane assuming the best fit value for $A_s$, as well as for its values in the wings of $68\%$
confidence limit. It shows that the allowed range is quite sensitive to the parameter $A_s$.
For smaller values of $A_s$ (but within its $68\%$ confidence limit) the flat model ($\Omega_k = 0$)
is more inconsistent with the data and negative curvature is preferred. But for higher values of $A_s$,
e.g $A_s= 0.81$ which still falls within the $68\%$ confidence limit, it shows that flat models as well
as models with small but both positive and negative curvature, are allowed [consistent with WMAP bound
on $\Omega_{total}$, $0.96 < \Omega_{total} < 1.08$ \cite{WMAP}] but for non zero values of $\alpha$. It suggests
that assuming a GCG model, one can alleviate the problem of consistency with a flat universe as pointed
out by \citeN{Choudhury2003b} for a $\Lambda$CDM model. 

In Figure 6, we have shown the same contours but now in the $\Omega_k$-$A_s$ plane assuming the best
values for $\alpha$ as well as its values at the wings of the $68\%$ confidence limit. Figure (6c)
is for $\alpha = 0.052$ which is almost a $\Lambda$CDM model, and it is quite similar to what obtained
by \citeN{Choudhury2003b} for a $\Lambda$CDM model. It shows that model that behaves more like $\Lambda$CDM
($\alpha = 0.052$) is inconsistent with a flat universe at $68\%$ confidence level. On the other hand, model
that deviates more from the $\Lambda$CDM model, Figure (6a) and (6b), is consistent with the flat universe
with better confidence level. Like Figure 5, it again shows that even if one does not take a flat prior,
unlike the $\Lambda$CDM model, flat GCG model is consistent with the supernova data
up to $68\%$ confidence level.

\section{Expected SNAP confidence regions}



\subsection{Method}

To find the expected precision of a future experiment such as SNAP, one must assume 
a fiducial model, and then simulate the experiment assuming it as a reference
model.This allows for estimates of the precision that the 
experiment might reach [see Silva \& Bertolami (2003) and references
therein for a more detailed 
description of the method employed here]. Let us then assume a fiducial
model and functions $\chi^2$ based on hypothetical
magnitude measurements at the various redshifts. In this case, 

\beq
\chi^2(\textrm{model})=\sum_{z_i=0}^{z_{max}}{\left[m_{\textrm{model}}(z_i)-m_{\textrm{fid}}(z_i)\right]^2 \over
\sigma^2(z)}~~,
\eeq
where the sum is made over all redshift bins and $m(z)$ is as specified in Section 3.
\begin{table}
\caption{SNAP specifications for a two year period of observations.}
\begin{center}
\label{SNAPerrors}
\begin{tabular}{lr}
\hline
\hline
Redshift Interval & Number of SNe\\
\hline
0.0-0.2............... & 50\\
0.2-1.2............... & 1800\\
1.2-1.4............... & 50\\
1.4-1.7............... & 15\\
\hline
\end{tabular}
\end{center}
\end{table}

\begin{figure}
\begin{center}
{\resizebox{0.45\textwidth}{!}{\includegraphics{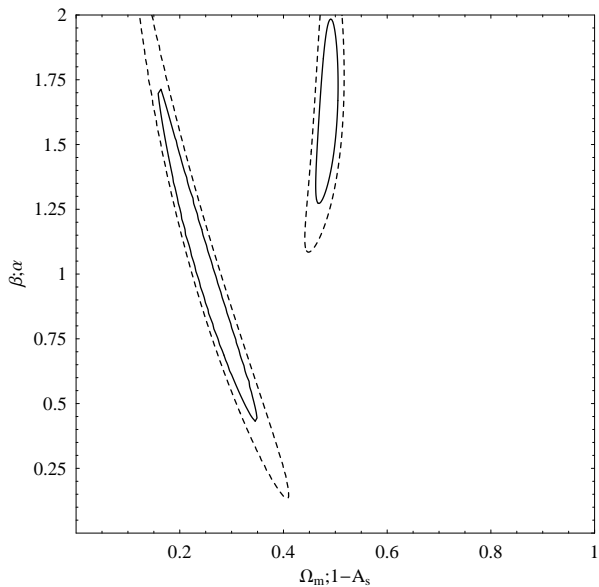}}}
\caption{Expected confidence regions for SNAP for a GCG fiducial
model with $1-A_s=0.25$ and $\alpha=1$. The solid and dashed lines represents
the $68\%$ and $95\%$ confidence regions respectively. The left contours are for GCG
and the right ones are for XCDM. For GCG, the parameter space is $1-A_s$ Vs. $\alpha$
whereas for XCDM, it is $\Omega_m$ Vs. $\beta$.
We have marginalized over ${\cal M}$.
}
\label{ChapFid}
\end{center}
\end{figure}

\begin{figure}
\begin{center}
{\resizebox{0.45\textwidth}{!}{\includegraphics{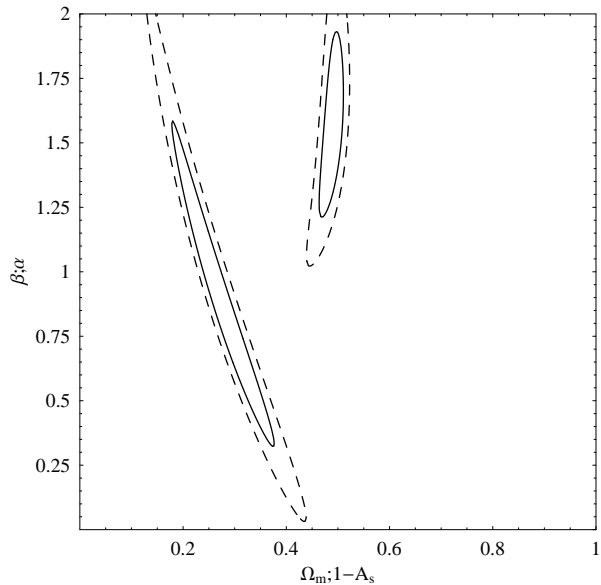}}}
\caption{Same as figure 7 but for a XCDM fiducial
model with $\Omega_m=0.49$ and $\beta=-\omega_x=1.55$.}
\label{DEFid}
\end{center}
\end{figure}

Here together with the GCG model we also analyze a flat XCDM model 
(that is a Cold dark matter model together with a dark energy with an equation
of state $ p_x = \omega_x \rho_x$).
We aim to show that if the Universe is indeed 
described by the GCG model, the fitting of a XCDM model to the data
will reveal that the cosmic expansion is drawn by a phantom-like dark energy component.

To fully determine the $\chi^2$ functions, the error estimates for SNAP
must be defined. Following \citeN{Albrecht}, we assume
that the systematic errors for the apparent magnitude, $m$, are given by

\beq
\sigma_{sys}={0.02 \over1.5}z ~~,
\eeq
which are measured in magnitudes such that at $z=1.5$ the systematic
error is $0.02$ mag, while the statistical errors for $m$ are estimated
to be $\sigma_{sta}=0.15$ mag. We place the supernovae in bins of width 
$\Delta z\approx0.05$. We add both kinds of errors quadratically

\beq
\sigma_{mag}(z_i)=\sqrt{\sigma^2_{sys}(z_i)+{\sigma^2_{sta} \over n_i}}~~,
\eeq
where $n_i$ is the number of supernovae in the $i'th$ redshift bin.
The distribution of supernovae in each redshift bin is, as before, taken from
\citeN{Albrecht}, and shown in Table I.

Summarizing, for each fiducial model, the method used, consists in the
following
\begin{enumerate}
\item{Choose a fiducial model.}
\item{Fit the XCDM model to the mock data, and obtain the respective
confidence regions.}
\item{Repeat the previous step to the GCG.}
\end{enumerate}

\begin{figure}
\begin{center}
{\resizebox{0.45\textwidth}{!}{\includegraphics{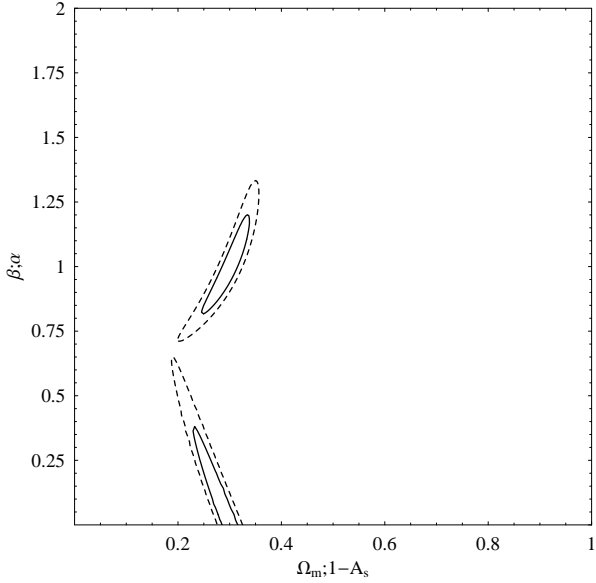}}}
\caption{Same as figure 7, but for a $\Lambda$CDM fiducial 
model with $\Omega_m=0.30$.}
\label{LCDMFid}
\end{center}
\end{figure}


\subsection{Results}

In Figures 7 to 9 we show the  confidence contours for the GCG and XCDM model 
for future SNAP observation taking different fiducial models. In all these Figures 
$\beta \equiv -\omega_x$. Also, along x-axis in all of these figures, we have plotted $1-A_s$ 
(instead of $A_s$ as $1 - A_s$ represents $\Omega_m$ when $\alpha = 0$) for the GCG and $\Omega_m$ for the XCDM. 
 
As mentioned above, our main aim is to explicitly show that a GCG Universe might appear
as a XCDM Universe with a dark energy component that has a 
phantom-type equation of state. To do so, we have considered 
two fiducial models. The first corresponds to a Chaplygin model ($\alpha=1$), with 
$1-A_s=0.25$. If one attempts to fit a XCDM model 
to the data (Figure \ref{ChapFid}), one finds that data favour a 
larger amount of matter than expected and a phantom-type dark
energy component. This is fully consistent with Figure 13 of
Tonry et al. (2003).

To further examine the degeneracy between models, in Figure
\ref{DEFid} we have repeated the procedure assuming a XCDM fiducial model, 
with $\Omega_m=0.49$ and $\beta=1.55$ ($w=-1.55$). From examining Figures
\ref{ChapFid} and \ref{DEFid}, one can see that both models appear
essentially identical. Also for GCG, the confidence regions for both  
fiducial models are quite identical to what we have shown earlier in Figure 2
for the current Supernova data. 

In Figure \ref{LCDMFid} we used a fiducial $\Lambda$CDM model, with
$\Omega_m=0.3$. As can be seen, the confidence regions are completely different from those found
in the two previous cases. Also, for the GCG model the confidence regions are quite different 
from what we have earlier in Figure 2 for the current Supernova data, further hinting that
indeed the $\Lambda$CDM model is not a  good description of the Universe.

To illustrate the degeneracy between the GCG model and the XCDM model with a phantom like equation of 
state ($\omega_x < -1$), we Taylor expand the luminosity distance as,

\beq
d_L = {c\over{H_0}}\left[ z + {(1-q_0)\over{2}}z^2 - {1\over6}(1-q_0-3q_0^2+j_0)z^3 + ....\right]
\eeq
where $q_0$ is the deceleration parameter related with the second derivative of the scale-factor and 
$j_0$ is the so-callled jerk parameter \cite{Visser2003}  related with the third derivative of the scale-factor. 
This is also one of the statefinder variables $r$ proposed in \citeN{Sahni}. The subscript ``0'' means that 
quantities are evaluated at present. 
The jerk parameter is related with the deceleration parameter $q_0$ as
\beq
j_0 = q_0 + 2q_0^2 + {dq\over{dz}}|_0 ~~.
\eeq
For the GCG model, one can calculate $q_0$ and ${dq\over{dz}}|_0$ to get
\begin{eqnarray}
q_0^{GCG} &=& {3\over{2}}(1-A_s) - 1 \nonumber\\
{dq\over{dz}}|_0^{GCG} &=& {9\over{2}}A_s(1-A_s)(1+\alpha),
\end{eqnarray}
whereas for XCDM model they turn out to be
\begin{eqnarray}
q_0^{DE} &=& {3\over{2}}(1-\omega_x(\Omega_m-1)) - 1 \nonumber\\
{dq\over{dz}}|_0^{DE} &=& {9\over{2}}\omega_x^2(1-\Omega_m)\Omega_m ~~.
\end{eqnarray}
For the previous Supernova data obtained by \citeN{Perlmutter1999} and \citeN{Riess1998} for low redshifts ($z < 1$), 
it is sufficient to consider the first two terms in the series expansion of the luminosity distance $d_L(z)$ given above. 
In that case, one can see from the expression of $q_0$ for GCG, that Supernova data can only constrain $A_s$, as $q_0$, for the 
GCG model, is independent of $\alpha$. Moreover, in order to have degeneracy between the GCG and XCDM model, 
the $q_0$ parameter of these two models must be equal, which results that:
\beq
A_s = \omega_x(\Omega_m - 1)~~.
\eeq
Thus, if one gets a bound on $A_s$ by fitting GCG model with low redshift Supernova data, 
then the same data can be fitted by a host of XCDM models (including $\Lambda$CDM model) provided the above 
equation is satisfied. Hence, for low redshifts ($z<1$), GCG model is degenerate with all kind of different 
dark energy models with constant equation of state, including the $\Lambda$CDM model.

Now, as one goes to higher redshifts, which is the case for the current data that we are considering in this paper, 
one also has to consider the higher order terms in the series expansion of $d_L(z)$. As far the data we are studying in our paper, 
it is enough to consider terms up to order $z^3$ in the series expansion of $d_L$. Hence, in order to 
have a degeneracy between GCG and XCDM even for the high redshifts, the jerk parameter $j_0$ also has to 
be equal for the two models, which effectively means ${dq\over{dz}}|_0$ has to be equal for the two models. 
Using this together with the equation (17), one finds that
\beqa
\omega_x &=& -\alpha(1-A_s)-1\nonumber\\
\Omega_m &=& {(1+\alpha)(1-A_s)\over{1+\alpha(1-A_s)}}~~.
\eeqa

In the above equation, $\omega_x$ and $\Omega_m$ are the equation of state and the 
density parameter of the  XCDM model which is degenerate with a GCG model with the 
corresponding $\alpha$ and $A_s$ parameters, for higher redshifts.  Now one can see 
that for any GCG model ($\alpha>0$), the corresponding equation state has to be always 
phantom type ($\omega_x < -1$) as $0<A_s<1$. This shows that although for the low redshift data, 
GCG model is degenerate with all kinds of constant equation of state dark energy model including 
the $\Lambda$CDM model, for higher redshift, the GCG is degenerate with a XCDM model but only with a phantom type of equation of state.

\begin{figure}
\begin{center}
{\resizebox{0.45\textwidth}{!}{\includegraphics{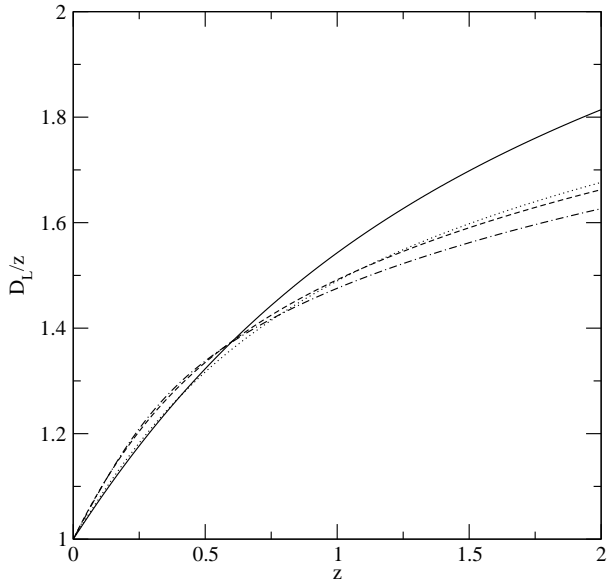}}}
\caption{Dimensionless luminosity distance $D_{L}/z$ as function of 
redshift for four different best fit models. The solid line is for $\Lambda$CDM. The dotted, dashed and dashed-dot 
lines are for Chaplygin, XCDM and GCG model, respectively. The values for the parameters of these models used in the 
plot are given in the main text.}
\label{figure10}
\end{center}
\end{figure}

Finally, aiming to further illustrate the degeneracy of the GCG model with the XCDM 
model for large red-shifts, we plot in Figure \ref{figure10} the behaviour of the dimensionless luminosity distance 
$D_{L}$ as function of 
redshift for four different best fit models, namely:
$\Lambda$CDM ($w_x = -1$, $\Omega_{m} = 0.3$), 
XCDM ($w_x = -2.07$ $\Omega_{m} = 0.51$), Chaplygin model ($\alpha = 1$, $A_s = 0.77$) and 
GCG ($\alpha = 3.75$, $A_s = 0.936$) respectively.
We have actually plotted the streched $D_{L}/z$ as function of redshift in order to graphically show the degeneracy.

\section{Conclusions}


We have analyzed the currently available 194 supernova data points within the
 framework of the
generalized Chaplygin gas model, regarding GCG as a unified 
candidate for dark matter and
dark energy. We have considered both, flat and non-flat cases, and used the 
best fit value for
${\cal M} = -0.033$  which is independent of a specific model, throughout our 
analysis.
For the first time, we have crossed the $\alpha = 1$ limit for the GCG model and
try to see whether the data actually allow it or not. 

For the flat case, we have studied both cases, restricting $\alpha$ to the 
range $0\leq\alpha\leq1$ and also without any restriction 
on the upper limit $\alpha$. From Figures 1 and 2, it is quite clear that data 
favours $\alpha > 1$,
although there is a strong degeneracy in $\alpha$. Also the
quality of fit improves substantially as one relaxes the $\alpha = 1$ 
restriction.
Moreover, the minimum values allowed for $\alpha$ and $A_s$ at $68\%$ 
confidence level are [0.78, 0.778], which excludes 
the $\alpha = 0$, $\Lambda$CDM case, although there is no 
constrain on $\alpha$ at $95\%$ confidence level.

Moreover, if one does not assume a flat prior for the analysis, our study shows that the
flat GCG model for $\alpha$ values sufficiently different from zero, is consistent with
the Supernova data up to $68\%$ confidence level. It also allows small, both positive and
negative curvature, making the GCG a somewhat better description than the $\Lambda$CDM model.
This is consistent with recent result which shows that without a flat prior, a flat
$\Lambda$CDM model, which is otherwise consistent with different cosmological observations,
 is not a good fit to the supernova data \cite{Choudhury2003b}. Moreover, the fact that GCG 
is a better fit to the Supernova data than $\Lambda$CDM, is consistent with the result of \citeN{Alam2003}, 
who also have reconstructed a similar kind of evolving equation of state for the dark energy 
from the latest Supernova data.

We have also studied the confidence contours for a GCG and XCDM model 
expected from the future SNAP observation assuming different fiducial 
universes.
In this regard, the degeneracy between the GCG model and a phantom-like
dark energy
scenario has made obvious in Section 4, where we have shown that 
when fitting a XCDM model to a GCG universe, the data will favour
a phantom energy component, and vice-versa. This degeneracy is also illustrated analytically through the expression for 
the luminosity distance $d_{L}$ as function of 
redshift. We have shown that for higher redshifts, GCG model is completely degenerate with a XCDM model with a phantom type of constant equation of state.
We mention that it has already been noted
in \citeN{Maor2002} that time varying equations of state might
be confused with phantom energy, and here we show that this is 
true for the GCG, without breaking the dominant energy condition
and without a big rip singularity in the future. It should also
be noted that with the exception of a cosmological constant,
most dark energy models predict a time varying equation of state,
therefore a constant dark energy equation of state might not
be the best parametrization for dark energy.
We have also shown that for a $\Lambda$CDM fiducial model, confidence regions for a GCG model, 
which are expected from future SNAP experiment, are quite different from what we have shown in Section 3.1, 
suggesting that SN Ia data does not favour the $\Lambda$CDM model.

Thus, our study shows that the generalized Chaplygin gas model is a very good fit to the latest Supernova 
data both with or without a flat prior. With future data, one expects the error bars to be reduced 
considerably, but we still expect that Supernova data will favour a generalized Chaplygin gas model with high confidence.

\vskip 2cm


\centerline{\bf {Acknowledgments}}

\vskip 0.2cm

\noindent
O.B.
acknowledges the partial support of Funda\c c\~ao para a 
Ci\^encia e a Tecnologia (Portugal)
under the grant POCTI/FIS/36285/2000. The work of A.A.S. is  
financed by  Funda\c c\~ao para a 
Ci\^encia e a Tecnologia (Portugal) under the grant SFRH/BPD/12365/2003. The work of S.S is financed by Funda\c c\~ao para a 
Ci\^encia e a Tecnologia (Portugal) through CAAUL.


\bibliography{mnrasmnemonic,astropap}

\begin{thebibliography}{}

\bibitem[\protect\citeauthoryear{{Alam} et al.}{{Alam} et al.}{2003}]
{Alam2003}Alam, U., Sahni, V., Saini, T. D., \& Starobinsky, A. A. 2003 
Preprint : astro-ph/0311364.

\bibitem[\protect\citeauthoryear{{Albrecht} \& {Skordis}}{{Albrecht}\& {Skordis}}{2000}]
{Albrecht2000}Albrecht, A., \& Skordis, C. 2000, \PRL 84, 2076

\bibitem[\protect\citeauthoryear{{Alcaniz} et al.}{{Alcaniz} et al.}{2002}]{Alcaniz}Alcaniz,
J. S., Jain, D., \& Dev, A. 2002, \PR D in press, Preprint : astro-ph/0210476

\bibitem[\protect\citeauthoryear{{Amendola}}{{Amendola}}{1999}]{Amendola1999}Amendola, L. 1999, \PR D60, 043501

\bibitem[\protect\citeauthoryear{{Avelino} et al.}{{Avelino} et al.}{2002}]{Avelino} Avelino, P. P.,
Be\c ca, L. M. G., de Carvalho, J. P. M., Martins, C. J. A. P., \& Pinto, P. 2002, \PR, D67, 023511

\bibitem[\protect\citeauthoryear{{Bachall} \& {Fan}}{{Bachall} \& {Fan}}{1998}]{Bahcall1998}Bahcall, N. A., \& Fan, X. 1998,
\AJ, 504, 1

\bibitem[\protect\citeauthoryear{{Balbi} et al.}{{Balbi} et al.}{2000}]{Balbi2000}Balbi, A., et al. 2000, \AJ 545, L1

\bibitem[\protect\citeauthoryear{{Banerjee} \& {Pav\'on}}{{Banerjee} \& {Pav\'on}}{2001a}]{Banerjee2001a}Banerjee,
N., \& Pav\'on, D. 2001a, \PR D63, 043504

\bibitem[\protect\citeauthoryear{{Banerjee} \& {Pav\'on}}{{Banerjee} \& {Pav\'on}}{2001b}]{Banerjee2001b}Banerjee,
N., \& Pav\'on, D. 2001b,  \CQG 18, 593

\bibitem[\protect\citeauthoryear{{Barris} et al.}{{Barris}}{2003}]{Barris2003}Barris, B. J., et al., 2003, Preprint : astro-ph/0310843

\bibitem[\protect\citeauthoryear{{Be\c ca} et al.}{{Be\c ca} et al.}{2003}]{Beca2003}Be\c ca, L.M.G., Avelino, P.P., de
Carvalho, J.P.M., Martins, C.J.A.P., preprint : astro-ph/0303564

\bibitem[\protect\citeauthoryear{{Benoit} et al.}{{Benoit} et al.}{2002}]{Archeops}Benoit, A., et al. 2002, \AAP, in press, Preprint : astro-ph/0210306

\bibitem[\protect\citeauthoryear{{Bento} \& {Bertolami}}{{Bento} \& {Bertolami}}{1999}]{Bento1999}Bento, M. C., \& Bertolami, O.
1999, \GRG 31, 1461

\bibitem[\protect\citeauthoryear{{Bento} et al.}{{Bento} et al.}{2001}]{Bento2001b}Bento, M. C., Bertolami,
O., \& Silva, P. T. 2001, \PL B498, 62

\bibitem[\protect\citeauthoryear{{Bento} et al.}{{Bento} et al.}{2002a}]{Bento2002a}Bento, M. C., Bertolami,
O., \& Santos, N. C. 2002a, \PR  D65, 067301

\bibitem[\protect\citeauthoryear{{Bento} et al.}{{Bento} et al.}{2002b}]{Bento2002b}Bento, M. C., Bertolami,
O., \& Sen, A. A. 2002b, \PR  D66, 043507

\bibitem[\protect\citeauthoryear{{Bento} et al.}{{Bento} et al.}{2003a}]{Bento2003a}Bento, M. C., Bertolami,
O., \& Sen, A. A. 2003a, \PR D67, 063003

\bibitem[\protect\citeauthoryear{{Bento} et al.}{{Bento} et al.}{2003b}]{Bento2003b}Bento, M. C., Bertolami, O., \& Sen, A. A. 2003b, \PL B575, 172.

\bibitem[\protect\citeauthoryear{{Bertolami}}{Bertolami}{1986a}]{Bertolami1986a}Bertolami, O. 1986a, Il Nuovo
Cimento, 93B, 36

\bibitem[\protect\citeauthoryear{{Bertolami}}{Bertolami}{1986b}]{Bertolami1986b} Bertolami, O. 1986b, \FP, 34, 829

\bibitem[\protect\citeauthoryear{{Bertolami} \& {Martins}}{{Bertolami} \& {Martins}}{2000}]{Bertolami2000}Bertolami,
O., \&  Martins, P. J. 2000, \PR D61, 064007

\bibitem[\protect\citeauthoryear{{Bili\'c} et al.}{{Bili\'c} et al.}{2002}]{Bilic2002}Bili\'c, N., Tupper, G.
B., \& Viollier, R. D. 2002, \PL B535, 17

\bibitem[\protect\citeauthoryear{{Bin\'etruy}}{{Bin\'etruy}}{1999}]{Binetruy1999}Bin\'etruy, P. 1999, \PR D60, 063502

\bibitem[\protect\citeauthoryear{{Bronstein}}{{Bronstein}}{1933}]{Bronstein1933}Bronstein, M. 1933, Phys. Zeit. Soweijt
Union, 3, 73

\bibitem[\protect\citeauthoryear{{Burles} et al.}{{Burles} et al.}{2001}]{Burles2001}Burles, S., Nollet, K. M.,
\& Turner, M. S. 2001, \AJ 552, L1

\bibitem[\protect\citeauthoryear{{Caldwell}}{{Caldwell}}{2002}]{Caldwell2002}Caldwell R. R., 2002, \PL B545, 23.

\bibitem[\protect\citeauthoryear{{Caldwell} et al.}{{Caldwell} et al}{1998}]
{Caldwell1998}Caldwell, R. R., Dave, R., \&  Steinhardt, P.J. 1998, \PRL 80, 1582

\bibitem[\protect\citeauthoryear{{Carlberg} et al.}{{Carlberg} et al.}{1998}]{Carlberg1998}Carlberg, R. G., Yee, H. K. C.,
Morris, S. L., Lin, H., Ellingson, E., Patton, D., Sawicki, M., \& Shepherd, C.
W. 1999, \AJ 516, 552

\bibitem[\protect\citeauthoryear{{Choudhury} \& {Padmanabhan}}{{Choudhury} \& {Padmanabhan}}{2003}]{Choudhury2003b}
Choudhury T.R., and Padmanabhan T., Preprint : astro-ph/0311622.

\bibitem[\protect\citeauthoryear{{de Bernardis} et al.}{{de Bernardis} et al.}{2000}]{deBernardis2000}de Bernarbis, P., et al.
2000, \NAT 404, 955

\bibitem[\protect\citeauthoryear{{de Bernardis} et al.}{{de Bernardis} et al.}{2002}]{Boomerang}de Bernardis, P., et al. 2002,
\AJ 564, 559

\bibitem[\protect\citeauthoryear{{Dev} et al.}{{Dev} et al.}{2002}]{Dev1}Dev, A., Alcaniz, J. S.,
\& Jain, D. 2002, \PR D67, 023515 

\bibitem[\protect\citeauthoryear{{Erickson} et al.}{{Erickson} et al.}{2002}]{Erickson2002}Erickson, J.K., Caldwell, R.R., Steinhardt, P.J.,  Armendariz-Picon, C., \& Mukhanov, V. 2002, \PRL, 88, 121301

\bibitem[\protect\citeauthoryear{{Ferreira} \& {Joyce}}{{Ferreira} \& {Joyce}}{1998}]{Ferreira1998}Ferreira, P. G.,
\& Joyce, M. 1998, \PR D58, 023503

\bibitem[\protect\citeauthoryear{{Fujii}}{{Fujii}}{2000}]{Fujii2000} Fujii, Y. 2000, \PR, D62, 064004

\bibitem[\protect\citeauthoryear{{Garnavich} et al.}{{Garnavich} et al.}{1998}]{Garnavich1998}Garnavich, P. M., et al. 1998,
\AJ, 509, 74

\bibitem[\protect\citeauthoryear{{Gong} \& {Duan}}{{Garnavich} \& {Duan}}{2004}]{Gong2004}Gong, Y \& Duan, C, Preprint : astro-ph/0401430

\bibitem[\protect\citeauthoryear{{Jaffe} et al.}{{Jaffe} et al.}{2001}]{Jaffe2001}Jaffe, A. H., et al. 2001, \PRL, 86,
3475

\bibitem[\protect\citeauthoryear{{Kamenshchik} et al.}{{Kamenshchik} et al.}
{2001}]{Kamenshchik2001} Kamenshchik, A., Moschella, U., \& Pasquier, V. 2001, \PL  B511, 265

\bibitem[\protect\citeauthoryear{{Kim}}{{Kim}}{1999}]{Kim1999}Kim, J. E. 1999, \JHEP, 05, 022

\bibitem[\protect\citeauthoryear{{Makler} et al.}{{Makler} et al.}{2002}]{Makler}Makler, M., Oliveira, S. Q.,
\& Waga, I. 2002, \PL B555, 1

\bibitem[\protect\citeauthoryear{{Maor} et al.}{{Maor} et al.}{2002}]{Maor2002}Maor, I., Brustein, R., McMahon, J.,
\& Steinhardt, P.J. 2002, \PR, D65, 123003

\bibitem[\protect\citeauthoryear{{Masiero} et al.}{{Masiero} et al.}{2000}]
{Masiero2000}Masiero, A., Pietroni, M., \& Rosati, F. 2000, \PR, D61, 023504

\bibitem[\protect\citeauthoryear{{Nesseris} \& {Perivolaropoulos}}{{Nesseris} \& {Perivolaropoulos}}{2004}]{Perivol2004}Nesseris, S \& Perivolaropoulos, L, 
Preprint : astro-ph/0401556

\bibitem[\protect\citeauthoryear{{Ozer} \& {Taha}}{{Ozer} \& {Taha}}{1987}]{Ozer1987}Ozer, M., \& Taha, M. O. 1987, \NP, 
B287, 776

\bibitem[\protect\citeauthoryear{{Padmanabhan} \& {Choudhury}}{{Padmanabhan} \& {Choudhury}}{2003}]
{Choudhury2003a} Padmanabhan T., Choudhury T.R., 2002, \MNRAS, 344, 823. 

\bibitem[\protect\citeauthoryear{{Peacock} et al.}{{Peacock} et al.}{2001}]{Peacock2001}Peacock, J. A., et al. 2001, \NAT,
410, 169

\bibitem[\protect\citeauthoryear{{Perlmutter} et al.}{{Perlmutter} et al.}{1999}]
{Perlmutter1999}Perlmutter, S., et al. 1999, \AJ, 517, 565

\bibitem[\protect\citeauthoryear{{Ratra} \& {Peebles}}{{Ratra} \& {Peebles}}{1988a}]
{Ratra1988a}Ratra, B., \& Peebles, P. J. E. 1988a, \AJ, 325, L117

\bibitem[\protect\citeauthoryear{{Ratra} \& {Peebles}}{{Ratra} \& {Peebles}}{1988b}]
{Ratra1988b}Ratra, B., \& Peebles, P. J. E. 1988b, \PR, D37, 3406

\bibitem[\protect\citeauthoryear{{Riess} et al.}{{Riess} et al.}{1998}]
{Riess1998}Riess, A. G., et al. 1998, \AJ, 116, 1009

\bibitem[\protect\citeauthoryear{{Sahni} et al.}{{Sahni} et al.}{2002}]
{Sahni} Sahni, V., Saini, T.D., Starobinsky, A.A., Alam, U., astro-ph/0201498.

\bibitem[\protect\citeauthoryear{{Sandvik} et al.}{{Sandvik} et al.}{2002}]
{Sandvik2002} Sandvik, H., Tegmark, M., Zaldarriaga, M., Waga, I., astro-ph/0212114

\bibitem[\protect\citeauthoryear{{Sen} \& {Sen}}{{Sen} \& {Sen}}{2001}]
{SenSen2001}Sen, A. A., \& Sen, S. 2001, \MPL A16, 1303

\bibitem[\protect\citeauthoryear{{Sen} et al.}{{Sen} et al.}{2001}]
{Sen2001}Sen, A. A., Sen, S., \& Sethi, S. 2001,\PR, D63, 107501

\bibitem[\protect\citeauthoryear{{Silva} \& {Bertolami}}{{Silva} \& {Bertolami}}{2003}]
{Silva2003}Silva, P.T., \& Bertolami, O. 2003, \AJ, 599, 829

\bibitem[\protect\citeauthoryear{{Spergel} et al.}{{Spergel} et al.}{2003}]
{WMAP}Spergel D.N. et al., 2003, \AJS, 148, 175

\bibitem[\protect\citeauthoryear{{Tonry} et al.}{{Tonry} et al.}{2003}]
{Tonry2003}Tonry J. L. et al., 2003, \AJ, 594, 1

\bibitem[\protect\citeauthoryear{{Turner}}{{Turner}}{2000}]{Turner2000}Turner, M. S. 2000, \PS, T85, 210

\bibitem[\protect\citeauthoryear{{Uzan}}{{Uzan}}{1999}]{Uzan1999}Uzan, J. P. 1999, \PR, D59, 123510

\bibitem[\protect\citeauthoryear{{Visser}}{{Visser}}{2003}]{Visser2003}Visser, M. gr-qc/0309109.


\bibitem[\protect\citeauthoryear{{Weller} \& {Albrecht}}{{Weller} \& {Albrecht}}{2002}]{Albrecht}Weller, J., \& Albrecht, A. 2002, \PR, D65, 103512

\bibitem[\protect\citeauthoryear{{Wetterich}}{{Wetterich}}{1988}]{Wetterich1988}Wetterich, C. 1988, \NC, B302, 668

\bibitem[\protect\citeauthoryear{{Zlatev} et al.}{{Zlatev} et al.}{1999}]{Zlatev1999}Zlatev, I., Wang, L.,
\& Steinhardt, P. K. 1999, \PRL, 82, 896

\end{thebibliography}
 
\bibliographystyle{mnras}

\end{document}